# Nanoscale Phenomena in Oxide Heterostructures

Joseph A. Sulpizio

*Department of Condensed Matter Physics, Weizmann Institute of Science, Rehovot 76100, Israel*

joseph.sulpizio@weizmann.ac.il

Shahal Ilani

*Department of Condensed Matter Physics, Weizmann Institute of Science, Rehovot 76100, Israel*

shahal.ilani@weizmann.ac.il

Patrick Irvin

*Department of Physics and Astronomy, University of Pittsburgh, Pittsburgh, PA 15260, USA*

prist2@pitt.edu,

Jeremy Levy

*Department of Physics and Astronomy, University of Pittsburgh, Pittsburgh, PA 15260, USA*

jlevy@pitt.edu

## Abstract
Recent advances in creating complex oxide heterostructures, interfaces formed between two different transition metal oxides, have heralded a new era of materials and physics research, enabling a uniquely diverse set of coexisting physical properties to be combined with an ever-increasing degree of experimental control. Already, these systems have exhibited such varied phenomena as superconductivity, magnetism, and ferroelasticity, all of which are gate-tunable, demonstrating their promise for fundamental discovery and technological innovation alike. To fully exploit this richness, it is necessary to understand and control the physics on the smallest scales, making the use of nanoscale probes essential. Using the prototypical LaAlO$_3$/SrTiO$_3$ interface as a guide, we explore the exciting developments in the physics of oxide-based heterostructures, with a focus on nanostructures and the nanoscale probes employed to unravel their complex behavior.

# Table of Contents

# 1. Introduction

For many decades, there has persisted a sharp dichotomy between the study of quantum materials and semiconductor/mesoscopic physics. The major thrust in the field of quantum materials has been the development and study of bulk, 3D correlated systems, where many physical phenomena can be achieved within a single materials family. A particularly fruitful family is that of the complex (transition-metal) oxides. Varying the composition and arrangement of the metal atoms in these oxides has enabled the synthesis of a diverse array of materials, including high-temperature superconductors, quantum magnets, ferroelectrics, and multiferroics. Meanwhile, the field of semiconductor physics has primarily focused on developing control over the dimensionality, cleanliness, and geometry of functional devices. Over the last ten years, these formerly independent communities have begun to overlap, beginning with the discovery of a high mobility two-dimensional electron gas living at the interface between strontium titanate ($SrTiO_3$) and lanthanum aluminate ($LaAlO_3$) (1). This emerging field of complex oxide heterostructures, devices made by combining different transition metal oxides, has enabled researchers to apply the experimental flexibility and tools of semiconductor physics to a family of materials that exhibits a multitude of intriguing physical properties.

The target audience for this review consists of researchers in mesoscopic physics, quantum transport, device engineering, and quantum information who are interested in this burgeoning field. Until recently, the main advances in this new field have been made by traditional materials science-based approaches. However, as complex oxide heterostructures reach an ever higher degree of maturity, researchers employing a broader class of approaches, from fundamental quantum physics to device applications, can find fertile ground for discovery.

To appreciate the potential novelty of oxide heterointerfaces, in Table 1 we highlight a selection of their properties as compared to those in other flagship solid state systems. Clearly, oxide interfaces are not yet as electronically clean as these other systems (although this is rapidly improving). However, their origin in strongly correlated materials endows these interfaces with a large set of properties that are rarely found to coexist in conventional materials systems. These include gate-tunable magnetism that persists to room temperature, intrinsic superconductivity, strong spin-orbit interactions, electron-lattice interactions, and possibly even ferroelectricity. Moreover, these properties have strong interrelationships, which once properly understood can be exploited to realize unique physical systems that have no counterpart in any other material system.

This review aims to provide an overview of the properties of oxide nanostructures, with an emphasis on the important role that local probes can play in gaining insight into this family of materials. Already, nanoscale probes have made important contributions to understanding and controlling the microscopic physics in the oxides. Scanning probes were critical in establishing the two-dimensionality of conducting electronic systems living near the interface (2), identifying the influence of microscopic structural domains on interfacial electrons (3, 4), and creating unique device architectures down to the nanoscale (5, 6). As will become clear over the course of this article, with so many different physical effects simultaneously at play in these systems, it is crucial to visualize and isolate the individual components on microscopic scales. Thus, nanoscale probes serve an increasingly vital role in the study of the oxides. The prototypical, and most widely-studied, oxide heterostructure is $LaAlO_3/SrTiO_3$, and so naturally this particular heterostructure will occupy much of the focus of this review. The knowledge gained, however, is endemic to the growing class of related oxide materials.

**Table 1: Comparison between LaAlO₃/SrTiO₃ and other solid state material systems for nanoscale devices.** Although LaAlO₃/SrTiO₃ is currently more electronically disordered than the other material systems, its broad array of physical properties and potential tunability make it very attractive as a system for studying correlated electron physics in engineered environments.

|  | LaAlO₃/SrTiO₃ | III-V Semiconductor heterostructures | Graphene | Nanotubes | Semiconducting Nanowires |
|---|---|---|---|---|---|
| **Dimensionality** | 2D / 1D | 2D / 1D | 2D | 1D | 1D |
| **Record Mobility ($cm^2/Vs$)** | >100,000 (40)* | 36,000,000 (150, 151) | 200,000 (152-154) | >100,000 (155) | 20,000 (156) |
| **Mean free path** | 100s of nm | ~100 μm | ~1 μm (153) | On substrate: ~10 μm (157) Suspended: not yet measured | ~100 nm (158) |
| **Typical densities** | $10^{12} - 10^{14}$ cm$^{-2}$ | $10^{10} - 10^{12}$ cm$^{-2}$ | $10^9 - 5 \cdot 10^{12}$ cm$^{-2}$ | $10^3 - 10^8$ cm$^{-1}$ | $10^5 - 10^8$ cm$^{-1}$ |
| **Achievable lithographic feature size** | few nm's with AFM lithography | ~100 nm (limited by interface depth (159)) | few nm's (but substantially increasing disorder) | tens of nm's (disorder-free in suspended devices (160)) | few nm's (grown heterointerfaces along the wires) |
| **Superconductivity** | yes $T_c^{max}$~300 mK gate-tunable | no | no (only by proximity) | no (only by proximity) | no (only by proximity) |
| **Magnetism** | yes $T_c$ > 300 K (20, 33) gate-tunable | yes $T_c$ < 200 K (161) | no | no | yes $T_c$ < 20 K |
| **Ferroelectricity** | Possibly | no | no | no | no |
| **Strong lattice-electron coupling** | Yes | no | no | yes | no |
| **Spin-Orbit coupling** | yes $few$ meV gate-tunable | yes ~1 meV gate-tunable Rashba & Dresselhaus | no $few$ μeV predicted | yes (162, 163) $few$ meV circumferential motion coupled to spin | yes ~200 μeV primarily Rashba |

\* γ-alumina top layer instead of LaAlO₃

## 2. SrTiO₃ – The universal substrate

Strontium titanate (SrTiO₃) serves as the near-universal substrate on which complex oxide structures are built and from which their properties are largely inherited. Thus, before proceeding to our discussions of nanoscale phenomena in complex oxide heterostructures, we first review the relevant physics of SrTiO₃, a material that has been extensively and continuously studied since the 1940s due to its remarkable array of properties. These diverse properties are best appreciated by considering how they emerge through the breaking of fundamental symmetries: point group symmetry (ferroelasticity), charge inversion symmetry (ferroelectricity), U(1) gauge symmetry (superconductivity), and spin rotation symmetry (ferromagnetism). We now discuss the specific breaking of each of these symmetries in the context of SrTiO₃, guided by the order in which they emerge as a function of decreasing temperature.

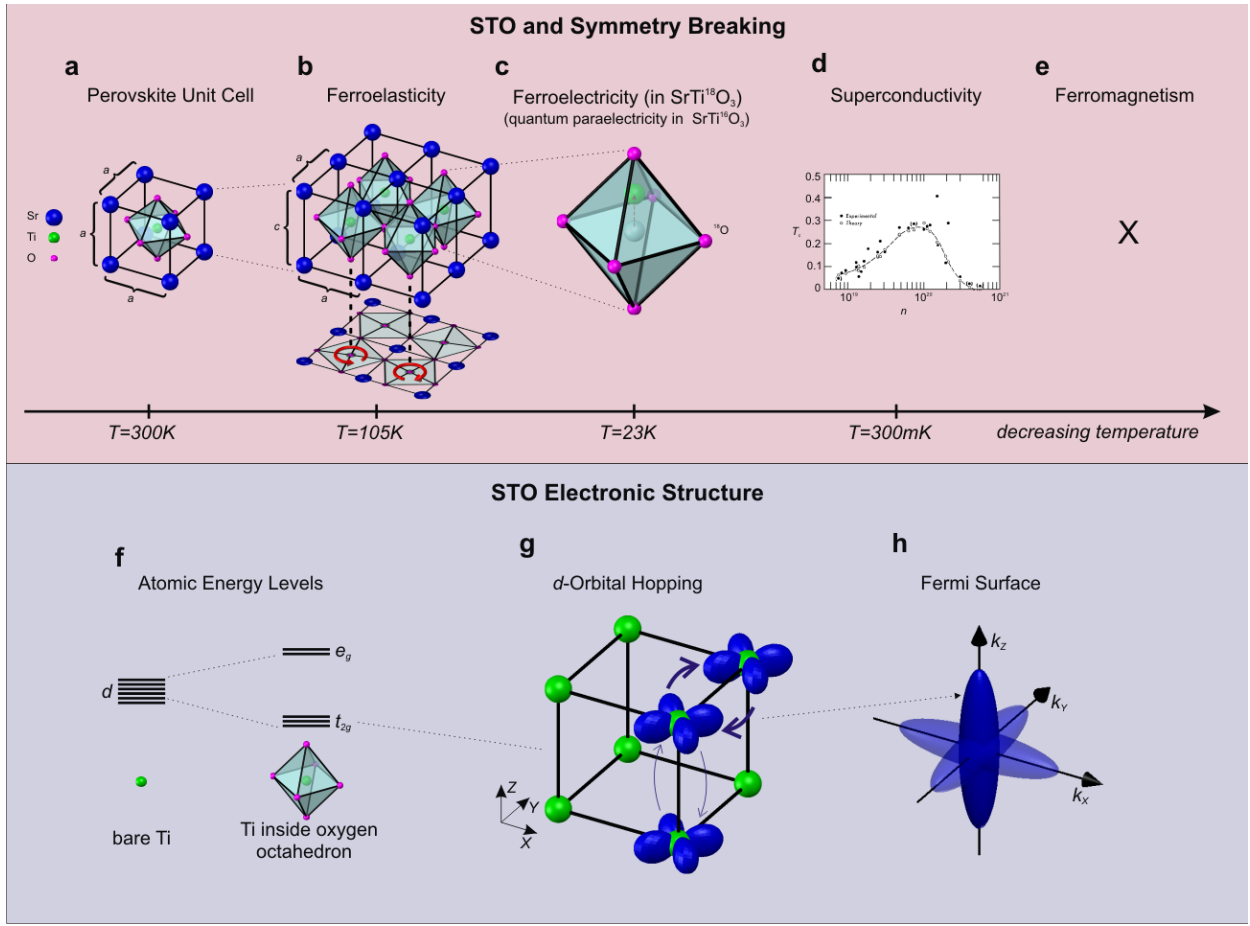

**Figure 1: SrTiO$_3$ - The universal substrate.** (a-e) The physical properties of SrTiO$_3$ emerge through the breaking of various symmetries as the temperature is lowered. (a) At room temperature, the Perovskite unit cell has cubic symmetry, with three orthogonal $a$-axes of equal length. (b) Below $T = 105$ K, a ferroelastic transition to tetragonal crystal symmetry occurs, where the crystal has two short $a$-axes and one long $c$-axis. This transition arises from an increasing "octahedral tilt", wherein neighboring oxygen octahedra develop a relative rotation (bottom projection, red arrows indicate rotation). (c) A ferroelectric transition occurs for SrTi$^{18}$O$_3$ at $T = 23$ K, endowing the unit cell with a permanent electric dipole from the displacement of the central titanium atom toward a corner of the distorted oxygen octahedron (dashed arrow). Quantum tunneling suppresses this transition for normal SrTi$^{16}$O$_3$, making it a "quantum paraelectric". (d) Near $T = 300$ mK, doped SrTiO$_3$ becomes a superconductor, (16). whose critical temperature exhibits a "dome" structure as a function of carrier density. (e) Rotation symmetry in spin space is always preserved, and therefore bulk SrTiO$_3$ does not exhibit magnetism at any temperature. (f-h) Electronic structure of SrTiO$_3$. (f) The electronic orbitals near the Fermi energy in SrTiO$_3$ are the fivefold-degenerate titanium $d$-orbitals. Once the titanium is surrounded by the oxygen octahedron, the $d$-orbitals are split into a higher energy doublet ($e_g$ states) and a lower energy triplet ($t_{2g}$ states). (g) Electrons located in the $t_{2g}$ orbitals ($d_{XY}$, $d_{YZ}$, $d_{XZ}$) are coupled to identical orbitals on neighboring lattice sites. Hopping matrix elements are much larger in the plane of an orbital's lobes than in the perpendicular direction. This is illustrated by the thickness of the arrows which indicates that hopping between $d_{XY}$ orbitals (blue) is stronger along the $X$ and $Y$ directions (light effective mass) than along the $Z$ direction (heavy effective mass). (h) The corresponding Fermi surface is cigar-shaped, elongated in the $k_z$ direction for the $d_{XY}$ orbitals (solid), and elongated in the $k_Y$ and $k_Z$ directions for the $d_{XZ}$ and $d_{YZ}$ orbitals (transparent), respectively.

## 2.1 Broken Symmetries in SrTiO$_3$

SrTiO$_3$ has a perovskite crystal structure, illustrated in Figure 1a. At room temperature its unit cell is cubic, composed of an outermost arrangement of eight strontium atoms. Centered on each face of the cube is one of six oxygen atoms, which form together the vertices of an octahedral cage. At the center of this cage lies the titanium atom. This cubic structure gives the most energetically favorable packing of the constituent atoms at high temperatures, but as the temperature is reduced, more efficient packing can be achieved. At $T = 105$ K, neighboring oxygen octahedra rotate in opposite directions (antiferrodistortive transition, Figure 1b) in order to reach the optimal bond lengths between both the strontium-oxygen and titanium-oxygen pairs (Goldschmidt tolerance) (7, 8). As a result of these rotations the cubic symmetry is broken and the unit cell becomes a rectangular prism with one long axis and two short axes (labeled c and a accordingly in Figure 1b). This ferroelastic transition naturally leads to the formation of domains within the SrTiO$_3$ with different orthogonal orientations of the tetragonal unit cells (9, 10). The consequences of this domain structure for nanoscale phenomena including the effect on local electronic properties of oxide heterostructures will be discussed in detail in subsequent sections. Further lowering of the crystal symmetry into orthorhombic and triclinic structures may occur at lower temperatures, though the magnitude of these additional perturbations is significantly smaller than the aforementioned transition to tetragonal symmetry. All such structural changes in the crystal symmetry lift the orbital degeneracies of the electronic system and thus have a direct effect on the electronic structure of the material.

After the cubic crystal symmetry of SrTiO$_3$ has been broken, the next symmetry to break is inversion symmetry. At room temperature the Ti atom is centered inside the oxygen cage and only through the application of an external electric field will it displace off-center, making SrTiO$_3$ a paraelectric material at these temperatures. This paraelectric state persists even below the ferroelastic transition. At even lower temperatures, however, the unit cell elongates enough such that a double-well potential develops for the titanium atom, with minima situated near opposite ends of the long axis of the oxygen octahedron (Figure 1c). For SrTi$^{18}$O$_3$, having the heavier isotope of oxygen, at $T = 23$ K a ferro*electric* transition occurs (11), in which the titanium atom spontaneously displaces toward one of these minima, resulting in a non-zero dipole moment within the unit cell. In contrast, the $^{16}$O isotope happens to be just light enough such that even at zero temperature it can quantum-mechanically tunnel between the two potential minima, a phenomena termed "quantum paraelectricity" (12). Similar to SrTi$^{18}$O$_3$, whose dielectric constant diverges at the ferroelectric transition, the dielectric constant of SrTi$^{16}$O$_3$ also rapidly increases with decreasing temperature, however, this increase is cut off by the quantum tunneling and this constant plateaus at a high value of ~20,000 (13), signifying the highly polarizable nature of this state. By growing thin films of SrTiO$_3$ on tensile-strained substrates, ferroelectricity can be observed as high as room temperature (14, 15).

Though SrTiO$_3$ is innately a band insulator, it can be made conducting by doping, for example with niobium. At very low temperatures ($T$~300 mK) doped SrTiO$_3$ becomes a superconductor (16) (Figure 1d), a property shared by many SrTiO$_3$-based oxide heterostructures. Finally, the breaking of rotational symmetry in spin space, which results in ferromagnetism, was never observed in bulk SrTiO$_3$ at any temperature. However, it was recently discovered (17) that ferromagnetism can emerge in SrTiO$_3$-based heterostructures, leading to pronounced magnetic ground states (17-33). Remarkably, as discussed in subsequent sections, these magnetic effects have been shown to coexist with superconductivity (19-21), two phenomena that are generally considered to be mutually exclusive. In fact, at these complex oxide interfaces, all four of the above mentioned symmetries have mutual interplay, leading to a very rich class of physical systems.

## 2.2 SrTiO$_3$ Electronic Structure

Having discussed the relevant symmetries, we now turn to the electronic structure of SrTiO$_3$, from which stems most of the physically observable quantities of complex oxide heterostructures. The energy states most relevant for the low energy physics of this system, those near the Fermi energy, arise from the fivefold degenerate $d$-orbitals of the Ti atom (Figure 1f). This degeneracy is lifted by the surrounding oxygen cage, which splits the levels into a high-energy doublet ($e_g$ states) and a low-energy triplet ($t_{2g}$ states) which remains near the Fermi energy. These $t_{2g}$ states comprise the $d_{XY}$, $d_{XZ}$, and $d_{YZ}$ orbitals, which couple to identical orbitals on Ti atoms at neighboring lattice sites through the $p$-orbitals of the oxygen atoms that lie in between. For a given orbital, the hopping matrix elements are much larger in the plane of the lobes than out of the plane. For example, for the $d_{XY}$ orbital the hopping is much stronger along the $X$ and $Y$ directions than along the $Z$ direction (Figure 1g), and as a result the effective mass is lower along the $X$ and $Y$ directions than it is along the $Z$ direction. In k-space this leads to a cigar-shaped Fermi surface along the $k_Z$ direction (Figure 1h). The $d_{XZ}$ and $d_{YZ}$ orbitals have analogous Fermi surfaces pointing along the $k_Y$ and $k_X$ directions, altogether forming a three-fold band structure centered at the Γ-point in k-space (34). The ferroelastic and ferroelectric distortions discussed above will lift the degeneracy of the three $t_{2g}$ orbital, as will spin-orbit interactions and the breaking of inversion symmetry near a surface or an interface, and this relation between crystal structure and electronic degeneracies creates interesting degrees of freedom which are still not fully characterized and utilized in these materials.

We note that the coexistence of such diverse phenomena in a single material can at first glance give the impression of a system that is hopelessly complex. However, the same degrees of freedom, if controlled, could lead to the creation of systems with great promise for technological applications and fundamental studies alike. While in the bulk these degrees of freedom can be controlled to a certain extent, the advent of oxide interfaces has allowed for the creation of systems that exploit the fundamental symmetries in entirely new ways, as we explore throughout the remainder of this review.

## 3. LaAlO$_3$/SrTiO$_3$ – A conducting layer between two insulators

While the discovery that conduction can dramatically emerge at the interface between a thin layer of LaAlO$_3$ grown on top of SrTiO$_3$ (Figure 2a) has led to a fury of research on oxide interfaces (35), there remain many fundamental questions about these systems. The most basic question, of why a conducting layer appears at all given that the two parent materials are insulating, continues to be an outstanding puzzle, and in this section we review the different mechanisms proposed to explain this phenomenon.

### 3.1 Influence of growth conditions

Whether an oxide interface is conductive or insulating is strongly dependent on the growth conditions, giving an important clue to the underlying mechanisms for conductivity. Oxide heterostructures, and specifically LaAlO$_3$/SrTiO$_3$, are primarily grown by Pulsed Laser Deposition (PLD), although Molecular Beam Epitaxy (MBE), Chemical Vapor Deposition (CVD) and sputtering are used as well. The most studied interface to date is that of crystalline LaAlO$_3$ grown over the (100) surface of SrTiO$_3$. Along this growth direction the SrTiO$_3$ is comprised of alternating layers of TiO$_2$ and SrO that are both charge neutral. LaAlO$_3$ grown along this direction is, on the other hand polar, consisting of alternating LaO and AlO$_2$ layers, which have charge +1 and -1 per unit cell, respectively. This results in a sharp transition at the interface between a non-polar and a polar material. Firstly, it was found that a conducting interface is achieved only when SrTiO$_3$ is terminated with a TiO$_2$ plane and not with an SrO plane (36). Moreover, the conductivity emerges abruptly only when the LaAlO$_3$ thickness reaches a critical value of four unit cells (37). LaAlO$_3$ stoichiometry also has an important influence on interface conductivity (38), as does oxygen partial pressure during growth and thermal annealing (37). Growth along the (111) direction which has a different polar discontinuity than (100), and the (110) direction that has no discontinuity at all have also exhibited conducting interfaces beyond a critical thickness, although in these cases it is not clear yet whether the critical thickness is universal as it is along the (100) direction. Non-polar overlayers such as amorphous LaAlO$_3$ (39) and spinel $\gamma$-alumina (40) also lead to interfacial conductivity, sometimes even with exceptionally high mobility (40). However, different from the LaAlO$_3$/SrTiO$_3$ case here the conductivity disappears with annealing in oxygen. Finally, even the surface of bare SrTiO$_3$ can be made to conduct, either by electrolytic top-gating of SrTiO$_3$ (41, 42) or by cleaving it in vacuum (43, 44).

For all of these interfaces it has been of primary interest to establish whether the conductivity occurs in the 3D bulk of SrTiO$_3$ or through a 2D layer near the interface (*i.e* a two-dimensional electron gas, or 2DEG). Here, nanoscale probes have played an essential role. Using conductive-tip atomic force microscopy the location of the conducting electrons was directly imaged in cross-sections of LaAlO$_3$/SrTiO$_3$ (2). These measurements revealed that conduction in samples grown under oxygen-poor conditions occurs throughout the bulk of the SrTiO$_3$; however, after oxygen annealing, it is concentrated on a narrow (<7 nm) region below the interface. Cross-sectional STM studies on LaAlO$_3$/SrTiO$_3$ have also measured a narrow 2D metallic region (~0.8 nm) (45), as have magnetotransport measurements, which are discussed in the next section.

## 3.2 Mechanisms for interfacial conduction

The above observations have led to conflicting views on the mechanisms generating conductivity at the interface, and this remains a hotly debated topic. Oxygen vacancies in SrTiO$_3$ (Figure 2b) can act as electron donors (46). The energetic barrier for the formation of these defects is quite low as compared for example to defects in semiconductors, and their formation could be further enhanced by fields near the interface (47). The observed sensitivity of interface conductivity to oxygen pressure during growth and to annealing supports this scenario. Cation intermixing across the interface in which atoms of different valence are exchanged may also play an important role (48, 49). For example, La atoms are known to readily dope bulk SrTiO$_3$ and thus their interchange with Sr atoms across the interface could lead to local doping and the generation of a conducting layer (Figure 2c). Transmission electron microscopy (36) and surface x-ray diffraction studies (50) have shown that the interface is not always atomically sharp, hinting that this scenario may be relevant.

There is however a fundamental intrinsic mechanism based on the presence of a strong polar discontinuity at the interface (51-53), often termed the "polar catastrophe" (36). For growth orientations that involve a polar discontinuity the potential difference across the LaAlO$_3$ increases linearly with its thickness, lifting the energy of its valance band with each additional monolayer (54-56). At a critical layer thickness, this band crosses the Fermi level and it then becomes energetically favorable to transfer electrons from the LaAlO$_3$ valance band edge at the surface to the bottom of the SrTiO$_3$ conduction band near the interface. The field generated by additional LaAlO$_3$ monolayers beyond this critical thickness would be compensated by further charge transfer from the surface to the interface, effectively pinning the surface LaAlO$_3$ valance band edge to the Fermi level. This charge transfer naturally results in the formation of a conducting 2D layer near the interface (Figure 2d). In contrast to the above mentioned scenarios which rely on local dopants that can be randomly distributed and are located close to the interface, the polar catastrophe scenario is reminiscent of field-effect gating, where charge is transferred from a spatially separated, homogenous region. One may imagine that conducting layers generated by this mechanism could be made free of disorder, assuming that one can eliminate other imperfections in the underlying lattice.

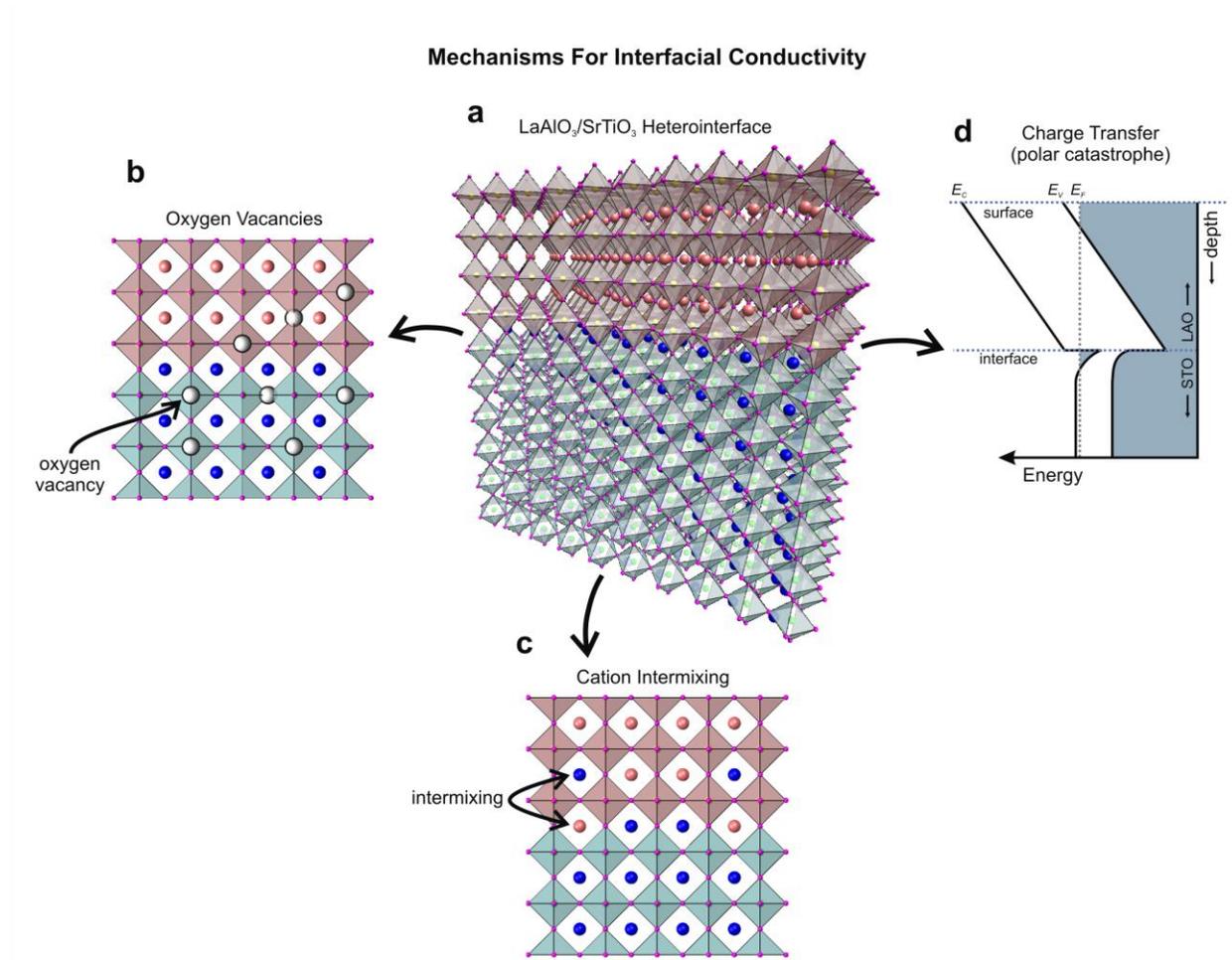

**Figure 2: Mechanisms for interfacial conductivity in oxide heterostructures.** (a) The LaAlO$_3$/SrTiO$_3$ heterointerface. Four monolayers of LaAlO$_3$ are shown atop a thicker SrTiO$_3$ substrate. The atoms are colored as: La (orange), Al (yellow), Sr (blue), Ti (green), O (purple). (b) Oxygen vacancies may form, donating charge to populate a conducting interface layer. (c) Cation intermixing, in which strontium atoms change position with lanthanum atoms across the interface will also effectively dope the interfacial layer. (d) Charge transfer/polar catastrophe: For polar growth orientations (e.g (100)), the built-in electric field in LaAlO$_3$ raises the energy of its bands with each layer. Beyond a critical LaAlO$_3$ thickness of 3 unit cells, the LaAlO$_3$ valence band edge of the top layer reaches the energy of the SrTiO$_3$ conduction band edge, and charge is transferred from the surface to the interface, populating a conducting interfacial layer. The energy bands are shown schematically. The Fermi level is marked by the vertical dashed gray line. The shaded regions represent occupied electron states, showing an excess of electron charge density in the SrTiO$_3$ conduction band near the LaAlO$_3$/SrTiO$_3$ interface donated from formerly occupied electron states at the surface, where an excess of holes now remains.

Strong support for the relevance of the polar catastrophe scenario comes from the fact that the observed critical thickness for conductivity in (100) LaAlO$_3$/SrTiO$_3$ is universal across many labs throughout the world employing a variety of growth techniques and conditions. However, it should be noted that the other mechanisms outlined above may also be sensitive to local polarization fields which may lead to very similar critical thickness dependence (47). The various scenarios described above are not mutually exclusive, and under generic growth conditions, multiple mechanisms may be at play. A systematic growth study (57) across the parameter space in which oxygen vacancy formation and charge transfer (polar catastrophe) are expected to be relevant indeed demonstrated how different mechanisms can coexist. With improved understanding and control over sample growth, it may be possible in the future to tailor the interface electrical properties based on a choice of a particular mechanism for the formation of the conducting layer.

In addition to controlling conducting interfaces via growth conditions, it is also possible to tune the conductivity *in-situ*, through a variety of approaches. Such tunability lends oxide heterostructures a great advantage over bulk oxides. Most prominently, the charge density of the two-dimensional conducting layer can be tuned through the electric field effect by the application of both back gate and top gate voltages (37, 58). It is also possible to reversibly change the interfacial conductivity using ferroelectric layers grown above the LaAlO$_3$. Applied to an LaAlO$_3$/SrTiO$_3$ sample just below the critical thickness, this approach (59) enables the switching between insulating and conducting phases. Polar adsorbates such as acetone and ethanol have also been shown to influence the conductive properties of LaAlO$_3$/SrTiO$_3$ heterostructures (60). Nanoscale patterning and scanned probe techniques designed to locally pattern the electrical properties of the conducting interface are discussed in the following sections.

## 4. Device fabrication

In order to realize functional electrical devices in oxide heterostructures, it is necessary to pattern the interfacial 2DEG. In conventional heterostructures such as GaAs-based quantum wells, the 2DEG is typically buried ~100 nm below the surface, which places harsh limits on the achievable device feature sizes. However, since in LaAlO$_3$/SrTiO$_3$ the 2DEG is only nanometers below the surface, the creation of devices with extreme nanoscale dimensions is possible. Here, we overview the main lithographic approaches to patterning LaAlO$_3$/SrTiO$_3$ down to the nanoscale: conventional lithography with LaAlO$_3$ thickness modulation, ion-beam irradiation, and surface modification including conductive-AFM lithography.

### 4.1 Lithography and LaAlO$_3$ thickness modulation

Conventional photo- and e-beam-lithographic techniques can be used to laterally define structures by locally controlling the thickness of the crystalline LaAlO$_3$ layer (61) (Figure 3a). In order to ensure that resist residue does not disrupt the conducting interface in the device region, 2 unit cells of LaAlO$_3$ are first deposited epitaxially over the entire substrate (TiO$_2$-terminated SrTiO$_3$). The sample is next removed from the growth chamber and patterned with photoresist after which amorphous LaAlO$_3$ is deposited and the remaining photoresist is lifted-off. After this process, the areas that were protected by resist contain pristine crystalline LaAlO$_3$, whereas the other regions are coated by amorphous LaAlO$_3$. Finally, epitaxial LaAlO$_3$ is grown to a thickness totaling >4 uc, defining the conducting regions (regions with amorphous LaAlO$_3$ on top remain insulating). Using this patterning method, conducting features have been made been as small as 1 μm with UV lithography and as small as 200 nm using e-beam lithography. A similar e-beam lithography technique using amorphous SrTiO$_3$ as a hard mask prior to growth of epitaxial LaAlO$_3$ has been used to fabricate devices as small as 500 nm (62).

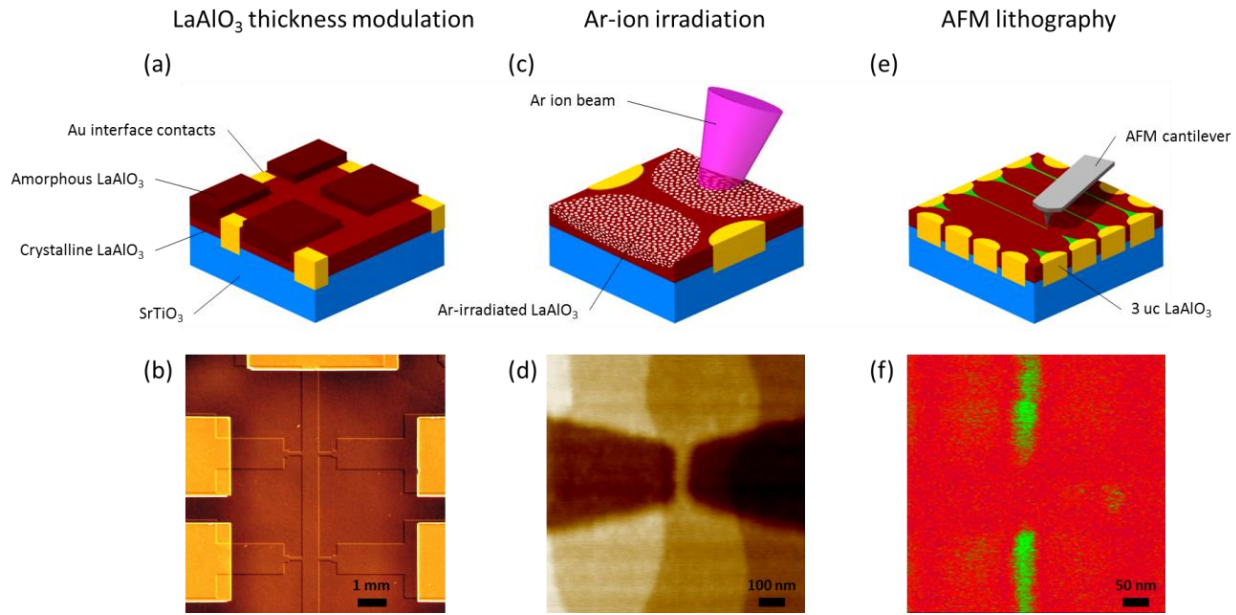

**Figure 3: LaAlO$_3$/SrTiO$_3$ device fabrication techniques.** Various fabrication techniques are illustrated schematically in the panels occupying the upper half of the figure, while the corresponding images of actual devices are shown in the lower panels. (a) Conventional lithographic techniques are used to form LaAlO$_3$ regions with thickness greater than 3 unit cells (conducting device region) surrounded by regions with 2 unit cells of LaAlO$_3$ (insulating). The latter are covered with amorphous LaAlO$_3$ (see text). Feature sizes down to 1 μm using UV photolithography and down to 200nm with e-beam lithography have been achieved using this technique. (b) Optical image of a Hall bar device with 10 μm features fabricated with this approach. (Optical image adapted with permission from Reference (61). Copyright 2006 AIP Publishing LLC.) (c) Argon ions can be used to transform conducting LaAlO$_3$/SrTiO$_3$ regions into insulating regions, most likely through the creation of local defects. (d) This technique can produce features as small as 50 nm, as shown by AFM imaging. (AFM image adapted with permission from Reference (63). Copyright 2013 AIP Publishing LLC.) (e) A conducting AFM probe is used to apply voltage to the surface of sub-critical thickness LaAlO$_3$ (3.3 unit cells), creating locally conducting regions with feature sizes below 10 nm. (f) Features created by c-AFM can be imaged using piezoelectric force microscopy. (Adapted with permission from Reference (68). Copyright 2013 AIP Publishing LLC.)

## 4.2 Ion beam irradiation

Fabrication of conducting structures down to 50 nm in size has been achieved using ion beam irradiation (63) (Figure 3c). In this process, a conducting LaAlO$_3$/SrTiO$_3$ (>4 uc) structure is first grown and e-beam lithography is used to define narrow regions in deposited resist that will ultimately remain conducting. Ar+ ions are then directed at the sample, and after this exposure, regions protected by resist remain conducting, while regions exposed to Ar+ become insulating, most likely due to the creation of local defects. Ion beams have also been used to modify the interface conductivity without the initial patterning of resist by using a stencil mask (64).

## 4.3 Conductive AFM lithography

Conducting features in LaAlO$_3$/SrTiO$_3$, down to just 2 nm, have been made using a conductive atomic force microscope (c-AFM) technique (Figure 3e) (65) to modify the charge state of the top LaAlO$_3$ surface. In doing so, critical thickness (~3 unit cell) LaAlO$_3$/SrTiO$_3$ interfaces can be locally and reversibly switched between conductive and insulating phases that are stable for hours in ambient conditions and indefinitely in vacuum or at cryogenic temperatures. The mechanism of formation for this metastable conductive state was found to be the local modification of the surface charge (66) through voltage-mediated addition and removal of water in the form of OH⁻ and H⁺ (67). Characterization via piezo force microscopy (68) (Figure 3f) and electric force microscopy (69) of c-AFM-written structures confirms their nanoscale size.

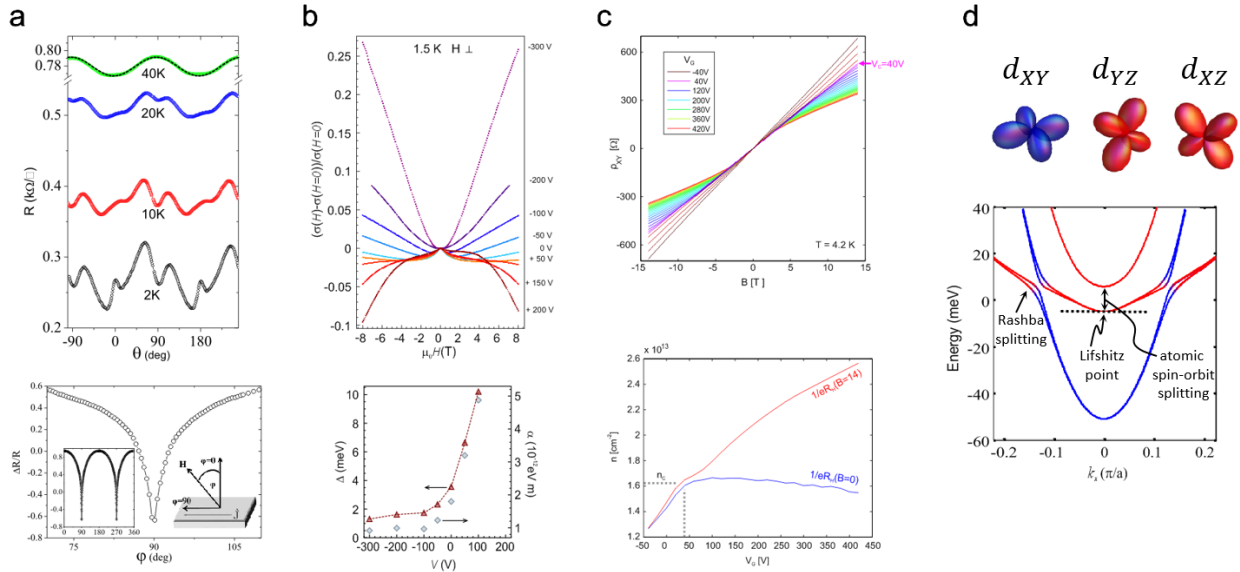

**Figure 4: Transport at the LaAlO$_3$/SrTiO$_3$ interface.** (a) Magnetoresistance anisotropy: Strong, negative magnetoresistance (lower panel) is observed, disappearing sharply as the magnetic field is tilted out of the plane. Strong anisotropies are also observed as a function of the angle of the field in the plane (upper panel) (Reproduced from (Reprinted with permission from Reference (80). Copyright 2009 by the American Physical Society.) (b) Spin-orbit coupling: Magnetoresistance curves change sign of field dependence as the gate voltage is made more positive (upper panel), from weak localization to weak anti-localization, indicating a strong onset of spin-orbit coupling, with characteristic scale reaching ~10 meV (lower panel). (Reprinted with permission from (82). Copyright 2010 by the American Physical Society.) (c) Lifshitz transition at a universal critical density: For high-mobility samples, a transition from linear Hall curves to non-linear Hall curves is observed, signifying a transition from single- to two-band transport. This transition occurs at a universal critical density $\sim 1.7 \times 10^{13}$ cm$^{-2}$ independent of electronic mobility and LaAlO$_3$ thickness. This critical density corresponds to a Lifshitz transition (lower panel), below which only a single, light band ($d_{XY}$) is occupied, and above which two additional heavy bands ($d_{YZ}$ and $d_{XZ}$) are populated. (Adapted by permission from Macmillan Publishers Ltd: Reference (85), copyright 2012.) (d) Schematic bands along the $k_x$ direction, with color-coded subbands matching the corresponding $d$-orbitals shown above.

## 5. Transport at the LaAlO$_3$/SrTiO$_3$ interface

The ability to confine electrons to a two-dimensional conducting layer in semiconductor heterostructures, mastered over the last four decades, has been a cornerstone in the creation of quantum systems with an extreme degree of tunability and ultra-high purity. The possibility to engineer analogous structures with transition-metal oxides creates an opportunity for even greater tunability due to the rich physics of the d-orbital electrons (70). Oxide-based systems are currently more disordered than III-V semiconductor systems, thus becoming conducting only at carrier densities that are two to three orders of magnitude higher (71) than in their semiconducting counterparts. As a result of this high doping, multiple electronic subbands are expected to be populated within the 2DEG, including carriers of widely different effective masses (light and heavy bands).

## 5.1 General transport phenomenology

To date, the main approach to studying the properties of LaAlO$_3$/SrTiO$_3$ interfaces has been through transport measurements (71-85). On the basic level, the angular dependence of the measured magnetotransport has established the two-dimensional nature of the conducting electron system (86). More generally, at high carrier densities (few $10^{13}$ cm$^{-2}$) the transport shows a rather complicated magnetic field dependence, attributed to the occupation of several bands (72-76). Nernst effect measurements (77) have also observed evidence for multiple bands. Several explanations have been proposed to account for the multi-band transport, and its characteristic gate voltage dependence. These models emphasized the importance of the confinement potential shape on the distance of the electronic wavefunctions from the interface and consequently the corresponding mobilities as set by the gate voltage (72, 78), as well as the effect of terraces in the underlying SrTiO$_3$ substrate (76, 79).

The occupation of many bands might suggest that the transport in LaAlO$_3$/SrTiO$_3$ should be quite complex and strongly dependent on sample-specific details such as disorder, preparation conditions, LaAlO$_3$ layer thickness, etc. Surprisingly, however, there are many unique transport phenomena, unfamiliar in e.g. III-V-based systems, universally found across LaAlO$_3$/SrTiO$_3$ samples created in various labs throughout the world. In the remainder of this section we highlight these commonly observed transport features, leaving discussions of superconductivity and magnetism for the following sections.

## 5.2 Anisotropic magnetoresistance

One of the most surprising observations is the unusual anisotropy of transport at strong magnetic fields. For fields oriented in the plane of the 2DEG, magnetoresistance is large and negative, often showing a dramatic (~6-fold) drop in resistance as compared to its value at zero field (80, 81). This negative magnetoresistance is very sensitive to the angle of the applied field and disappears when it is tilted slightly out of the plane (~1°, Figure 4a, lower panel). Furthermore, there is a strong anisotropy in transport with respect to the angle of the applied field within the plane of the 2DEG (Figure 4a, upper panel) (80). At very low temperatures, Shubnikov-de Haas oscillations in the longitudinal resistivity have also been observed (74) (73), reflecting a carrier density that is an order of magnitude lower than that measured by the Hall effect, a discrepancy that so far eluded a clear explanation. At small perpendicular fields, the magnetoresistance changes sign as a function of gate voltage from negative (weak localization) to positive (weak anti-localization (82, 83), Figure 4b). The latter is interpreted as arising from spin-orbit coupling with a rather large energy scale (~10 meV), comparable to the Fermi energy of the system. Curiously, the onset of spin-orbit effects (82, 84) occurs sharply at a certain carrier density, and is correlated rather well with the emergence of superconductivity.

## 5.3 A universal Lifshitz transition from single-band to multi-band transport

At sufficiently low carrier densities in high mobility samples, a simpler, more general view of transport begins to emerge. Experiments have shown (85) that in these samples there is a critical carrier density below which the transport follows a simple single-band behavior, evident from a Hall voltage that is linear in field. Above this density, a transition is observed to non-linear Hall curves, consistent with two band transport (Figure 4c). Curiously, this transition occurs at a universal critical density $\sim 1.7 \times 10^{13}$ cm$^{-2}$ across a broad range of sample mobilities and LaAlO$_3$ thicknesses (85). This universality suggests that the observed transition is different from a disorder-driven metal-insulator transition (71) and that it instead originates from intrinsic properties of the 2DEG.

The emerging picture is that the critical density corresponds to a Lifshitz transition between the population of only a single, light, circular $d_{XY}$ band, and the additional population of two heavy $d_{XZ}$ and $d_{YZ}$ bands (Figure 4d). This simple view was indeed recently confirmed by gate-dependent ARPES measurements (87). The sudden appearance of spin-orbit interactions as a function of gate voltage naturally follows from this band picture, since atomic spin-orbit interactions should be most prominent where bands are degenerate, which is precisely the situation of the heavy bands at the Lifshitz transition. Since the Rashba spin-orbit coupling is proportional to the atomic spin-orbit coupling, its influence will also be peaked at the Lifshitz transition (85). Several theoretical works have relied on this Rashba coupling to explain various magnetotransport phenomena in LaAlO$_3$/SrTiO$_3$ (83, 88, 89).

While the understanding of the basic elements of LaAlO$_3$/SrTiO$_3$ transport is beginning to coalesce, many important topics remain unresolved and not fully explored. Of particular note are the influence of disorder and sample inhomogeneities, the lifting of orbital degeneracies by the various symmetry breakings happening in SrTiO$_3$ (e.g. through local lattice distortions) and their role in transport, and how these factors can be controlled on the nanoscale with an eye toward novel device applications.

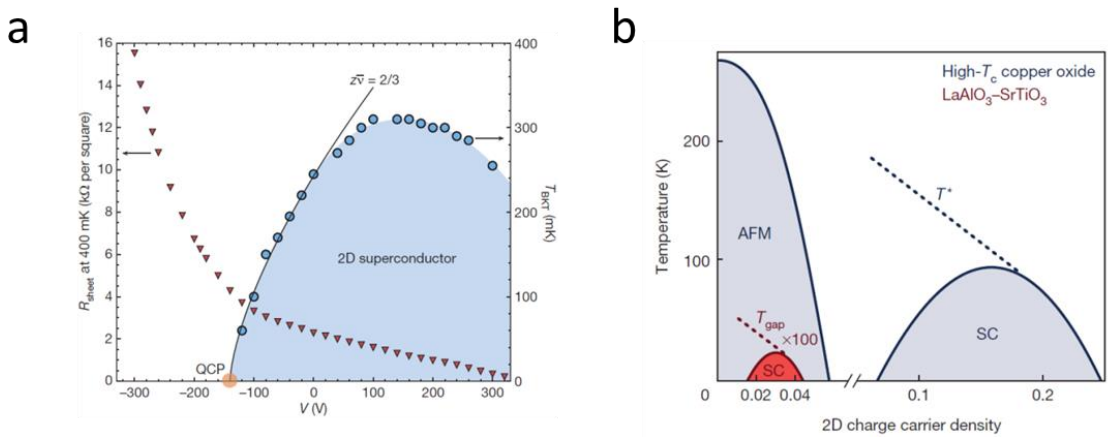

**Figure 5: Superconductivity.** (a) Superconducting critical temperature of LaAlO$_3$/SrTiO$_3$ (blue circles) and normal-state sheet resistance (red triangles) measured as a function of gate voltage. The superconductivity has a clear dome structure. (Adapted by permission from Macmillan Publishers Ltd: Reference (98), copyright 2008.) (b) Illustration of data from tunneling measurements that identify a characteristic line above the critical temperature at which a pseudo-gap opens (dashed red). This is strongly reminiscent of the pseudo-gap state in high-$T_c$ superconductors (drawn to scale in blue). (Adapted by permission from Macmillan Publishers Ltd: Reference (99), copyright 2013.)

## 6. Superconductivity

SrTiO$_3$ is the most dilute superconductor found in nature (90), and served as inspiration for Bednorz and Müller in their quest for high-temperature superconductors (91). Bulk SrTiO$_3$ becomes a superconductor (16, 92) at sufficiently high doping ($\sim 5 \times 10^{17}$ cm$^{-3}$) and sufficiently low temperatures ($T_c \sim 300$ mK). The superconducting critical temperature was shown to have a dome structure as a function of the 3D carrier density (92) (Figure 1d), reaching a maximum $T_c \sim 300$ mK for densities of $\sim 10^{20}$ cm$^{-3}$. While superconductivity in SrTiO$_3$ is generally believed to be phonon-mediated (BCS) (93, 94), there is still an ongoing debate on whether this superconductivity is unconventional, having a number of energy bands contributing to the superconducting state (95, 96), and possibly a non-standard pairing due to a significant role of spin-orbit interactions (95, 96).

## 6.1 Superconducting phase diagram

The 2DEG in LaAlO$_3$/SrTiO$_3$ heterostructures also becomes superconducting at sufficiently low temperatures (97), with features that are strongly reminiscent of bulk SrTiO$_3$ superconductivity. Again, a characteristic superconducting dome has been observed as a function of carrier density (Figure 5a) (98), with maximum $T_c \sim 300$ mK, close in value to that observed in the bulk. Due to this high degree of similarity with bulk superconductivity, it has been postulated that the superconductivity in LaAlO$_3$/SrTiO$_3$ can be understood by considering the effective 3D density that corresponds to the electrons localized near the interface. On the other hand, the superconducting dome also bears strong resemblance to the dome structure seen in high-$T_c$ superconductors (Figure 5b), and recent tunneling measurements have even observed the existence of a pseudogap in LaAlO$_3$/SrTiO$_3$ with similar carrier dependence to that observed in high-$T_c$ superconductors (99), suggesting this feature may be endemic to 2D superconductivity.

## 6.2 Local structure of superconductivity

Scanning SQUID measurements have provided insights into the structure of superconductivity on microscopic scales (19). These studies have shown that in LaAlO$_3$/SrTiO$_3$ superconductivity is spatially inhomogeneous, in contrast to superconductivity in delta-doped SrTiO$_3$. Additionally, the superfluid density in LaAlO$_3$/SrTiO$_3$ was shown to be tunable with gate voltage (100) and quite small in magnitude (few $10^{12}$ cm$^{-2}$) (98), ten times lower than the total carrier density measured via the Hall effect. Gate voltage dependence studies have also revealed a correlation between the peak of the superconducting dome and the Lifshitz transition between light and heavy subbands (85), emphasizing the possible importance of the different subband symmetries in the formation of the superconducting ground state.

## 6.3 Superconductivity and magnetic fields

Measurements of the superconducting critical magnetic field, $H_c$, demonstrate strong anisotropies for fields applied in and out of plane. Such anisotropy confirms that the superconductivity is two-dimensional in nature, placing an upper bound of 10-50 nm on the thickness of the superconducting layer (84, 101). Measurements of the critical perpendicular field, analyzed within the Ginzburg-Landau theory, give an estimated coherence length of $\sim 50 - 100$ nm. The critical in-plane field is curiously observed to be substantially larger than the Chandrasekhar-Clogston limit (up to 4x), which states that the conventional superconducting singlet is destroyed once the applied field is strong enough to parallelly align the two constituent spins, *i.e. $g\mu_B B \sim \Delta$*, where $g$ is the lande g-factor, $\mu_B$ is the Bohr magneton, $B$ is the magnitude of the in-plane field, and $\Delta$ is the superconducting energy gap. In-plane fields can even be exploited to increase the critical superconducting temperature (102). These observations highlight the strong effects of spin-orbit interactions and might suggest that an unconventional form of superconducting pairing may be at play.

Remarkably, as will be more fully discussed in the following section, superconductivity in LaAlO$_3$/SrTiO$_3$ is often observed to coexist alongside magnetism within the same sample (19). These phenomena are generally considered to be mutually exclusive, and it is not yet clear whether they occur in regions of the sample separated in space, or whether they occur within the same region, in which case the superconductivity would be of an unconventional form, possibly involving an important role of spin-orbit coupling (103, 104). Future investigations with nanoscale probes are necessary to resolve this. Recent application of AFM-based techniques has even observed superconductivity in quasi one-dimensional nanoscale structures (105). These observations pave the way to controlling superconductivity over nanometer length scales.

## 7. Magnetism

The existence of magnetism at the LaAlO$_3$/SrTiO$_3$ interface is perhaps one of the most surprising and least understood phenomena in this system. While neither of the parent compounds (SrTiO$_3$ and LaAlO$_3$) are magnetic, after an interface between them is formed, clear signatures of magnetic order emerge, often persisting up to room temperature. In this section we review the understanding gained by recent experimental and theoretical works on magnetism in LaAlO$_3$/SrTiO$_3$, and address the following fundamental open questions: is the magnetism attributable to localized electrons, itinerant electrons, or a combination of both? What are the magnetic ground states of the system? Can these magnetic states be tuned with gate voltage as can the electrical and mechanical properties of this interface?

### 7.1 Experimental manifestations of magnetism at the interface

So far, experiments uncovered two manifestations of magnetism. The first is ferromagnetism (FM), observed typically at low magnetic fields, involving a large number of spins and occurring in inhomogeneous patches that vary strongly between and within samples, and shows no gate voltage dependence. The second is seen via magnetoresistance measurements, and points to a gate voltage-dependent metamagnetic transition (rapid increase in magnetism at a finite applied magnetic field) that strongly involves the conduction electrons. We will discuss both of these manifestations below.

Signatures of FM in LaAlO$_3$/SrTiO$_3$ were first uncovered (17) through low temperature (300 mK) measurements of the 2DEG resistance that showed clear hysteresis as a function of magnetic field (Figure 6a), hinting at the existence of a field-switchable spontaneous magnetization. Soon thereafter FM hysteresis was demonstrated up to room temperature, and was shown to coexist with paramagnetic and diamagnetic susceptibilities (20), suggesting that the electrons have multiple coexisting phases. Similar hysteresis observed in the superconducting critical current (21) signified that the coexistence is with superconductivity. The existence of FM was further corroborated by torque magnetometry (22), which measured the magnetization of LaAlO$_3$/SrTiO$_3$ directly and showed that it is large even within the superconducting state (Figure 6b). The microscopic nature of this coexistence was elucidated by scanning squid experiments (19). These studies revealed that FM is concentrated in sparse random patches, each having a large magnetic dipole surrounded by a majority of non-FM superconducting regions (Figure 6c). To date, it is not known whether FM and superconductivity (or even normal conductivity) occur in the same regions within a sample, in different layers, or side by side. Within the FM patches the magnetic moment are aligned in-plane and their density is quite high (few $10^{14}$ cm$^{-2}$) (19, 23). However, as these patches are sparsely distributed, the overall interfacial moment density is rather low (few $10^{12}$ cm$^{-2}$) (19, 23). Such low density is consistent with measurements by various macroscopic probes (24-26), with the exception of torque magnetometry (22) experiments, which observed a moment density approaching the limit of half an electron per unit cell over the entire sample. X-ray circular dichroism has pinned down that the moments contributing to the ferromagnetism arise from localized $d_{XY}$ states of Ti atoms, which lie very close to the interface, most probably within the first TiO layer (26).

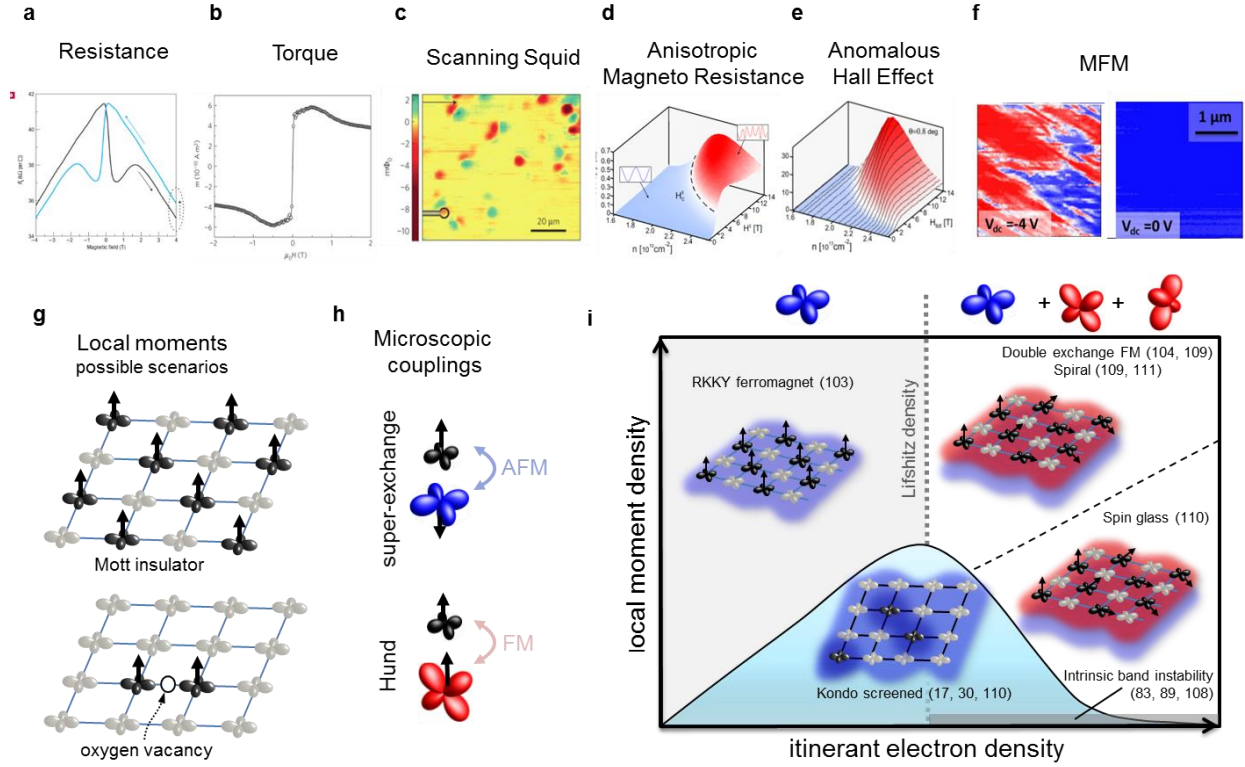

**Figure 6: Magnetism.** (a-f) Survey of various measurements of magnetism at the LaAlO$_3$/SrTiO$_3$ interface. The rich phenomenology includes observations of (a) hysteretic magnetoresistance, (Adapted by permission from Macmillan Publishers Ltd: Reference (17), copyright (2007).) (b) Large in-plane magnetization measured by torque magnetometry (Adapted by permission from Macmillan Publishers Ltd: Reference (22), copyright (2011).) (c) magnetic dipoles imaged with scanning squid microscopy (Adapted by permission from Macmillan Publishers Ltd: Reference (19), copyright (2011).) (d) a gate-tunable meta-magnetic transition between phases with weak (blue) and strong (red) anisotropic magnetoresistance (30), and (e) a corresponding emergence of an anomalous Hall effect (30), and (f) room temperature ferromagnetism measured via magnetic force microscopy (33). (g) The two proposed scenarios for the emergence of local magnetic moments in LaAlO$_3$/SrTiO$_3$. Top panel: A Mott insulator formed from a quarter-filled ($1/2\,e^-$ per unit cell) $d_{XY}$ energy band, localized very close to the interface. The black orbitals represent localized $d_{XY}$ states, and the arrows represent their local magnetic moments. Bottom panel: An alternative scenario in which the magnetic moments arise near oxygen vacancies. (h) Microscopic couplings between itinerant and localized moments. An effective antiferromagnetic (super-exchange) interaction due to Pauli exclusion couples itinerant $d_{XY}$ electrons (blue) to localized $d_{XY}$ moments (black). Conversely, for itinerant $d_{YZ}$ and $d_{XZ}$ electrons (red) hopping is allowed for both spin orientations, resulting in a ferromagnetic Hund's coupling to the localized $d_{XY}$ moments (black). (i) The emerging phase diagram, showing the magnetic phases formed in the plane parameterized by itinerant electron density and local moment density. Relevant itinerant orbitals for different density regimes are shown above the diagram. The critical Lifshitz density, marked by the dashed vertical line, indicates the density above which itinerant $d_{YZ}$ and $d_{XZ}$ electrons are populated.

## 7.2 Origin of ferromagnetic moments

Two fundamentally different scenarios are proposed for the emergence of FM moments in LaAlO$_3$/SrTiO$_3$. The first scenario (Figure 6g, upper panel) attributes them to an intrinsic instability of a special $d_{XY}$ energy band, theoretically predicted to split-off to lower energies and to be much more localized than the other interfacial bands, residing only in the first TiO plane adjacent to the interface (27). This band is expected to accommodate the majority of carriers transferred to the interface, which within the polarization catastrophe hypothesis is predicted to reach half an electron per Ti atom for thick LaAlO$_3$. Such a high carrier population effectively makes this band a quarter-filled Mott insulator, driving carriers to localize on every second Ti site, such that they don't participate in conduction, thereby explaining why only a small fraction of the total charge density is actually measured in transport. These localized carriers are extremely susceptible to magnetic polarization, either via a Stoner-like mechanism (27), or through interactions with even a small fraction of conduction electrons (103). The experimental observation (23) that FM emerges only above a critical LaAlO$_3$ thickness that is nearly equal to that in which the 2DEG conductivity emerges would be consistent with this scenario.

A second scenario (Figure 6g, lower panel) relates the FM moments to an extrinsic source. The observation of randomly distributed FM patches within a sample (19), the large variability between samples (19, 23), high sensitivity to strain (28), lack of gate-voltage dependence (19), and most importantly the emergence of FM only in samples grown at high oxygen pressures (17, 20) make oxygen vacancies and their possible aggregation within the sample a prime suspect. Indeed, DFT calculations show that Ti atoms around an oxygen vacancy should develop a large local magnetic moment (29). In contrast to the intrinsic scenario above, here the density of free moments will be highly variable and dependent on details of sample preparation.

In those samples that contain FM patches the magnetization within the patches is quite large. One may wonder whether in these samples magnetism also exists within the majority of 2DEG outside of these patches or even in samples that don't contain patches at all. It is reasonable to expect that a much smaller but still non-zero density of moments will exist in these regions, which could have density comparable to that of the conduction electrons. In this regime, the interaction between these dilute moments and conduction electrons could in fact give rise to interesting magnetic ground states in which these two components are players on equal footing.

Evidence for such magnetic ground states that strongly depended on the density of the itinerant electrons has been found in magnetotransport experiments (30) which indicate the existence of a metamagnetic transition beyond a critical magnetic field (c.f. Section 5). Around this field, dramatic changes are observed in the ground state of the system - below the critical field the magnetoresistance is isotropic in the plane of the interface, but directly above this field it becomes strongly anisotropic (~50%), with preferred axes along crystalline directions (30, 31) (Figure 6d). Concurrent with these observations is a strong anomalous Hall effect component that emerges when the critical field is crossed and saturates soon thereafter (30, 32), indicating a sudden buildup of internal magnetization (Figure 6e). Finally, the longitudinal resistance drops dramatically above critical field and saturates on a value that is smaller by almost an order of magnitude (30, 31, 80). These three effects occur in perfect synchrony, and are suggestive of a metamagnetic transition between a non-magnetic phase below the critical field and a polarized magnetic phase with preferred easy axes above it, the latter being similar to that observed in dilute magnetic semiconductors (106, 107). The metamagnetic critical field depends strongly on gate voltage, with its value smoothly increasing with decreasing carrier density and appearing to diverge as the Lifshitz point is approached from above (at the bottom of the heavy $d_{XZ}/d_{YZ}$ bands). This dependence indicates that the conduction electrons are playing a central role in the observed magnetic phases. The correlation between the divergence of the critical field and the Lifshitz point, which corresponds to the transition between solely populating the $d_{XY}$ band and additional population of the $d_{XZ}/d_{YZ}$ bands, signifies that the orbital symmetry of the conducting electrons have also an important effect on the resulting magnetic ground states. Gate-tunable magnetic states have further been observed at the low-density side of the phase diagram. Using MFM measurements it was discovered (Figure 6f) (33) that as the conduction electrons are depleted, the interface becomes FM even at room temperature, with domain structure on the micrometer scale. Unlike the FM discussed above, here the FM disappears completely as the conduction electrons are populated under the application of a gate voltage.

## 7.3 Magnetic phase diagram

Several theoretical works have studied the different magnetic phases that could appear at this interface. All these works can be cast into a generic phase diagram describing the magnetic order as a function of the itinerant carrier density and the local moment density (Figure 6i). One class of theories (83, 89, 108) assumed that impurities are irrelevant, explaining the observed magnetic order as an intrinsic Stoner-like instability of the itinerant $d$-bands. Such an instability could occur near the Lifshitz point, where the strong spin-orbit interactions that exist as the two heavy $d_{XZ}/d_{YZ}$ bands become populated could cause a large enhancement of the density of states. Other theories (30, 103, 104, 109-111) have emphasized the crucial role played by localized moments and their coupling to the itinerant electrons in determining the magnetic ground states. There exist two opposite microscopic couplings between the localized and itinerant electrons which follow directly from the symmetries of the underlying orbitals (Fig. 7h). Both localized and itinerant electrons originate from the Ti $d$-orbitals, the former populating only $d_{XY}$ states and the latter populating $d_{XY}$ states below the Lifshitz transition and additional $d_{XZ}/d_{YZ}$ states above this transition. Due to Pauli exclusion, an itinerant $d_{XY}$ electron can hop onto a Ti site that is occupied by a localized $d_{XY}$ moment only if it has the opposite spin, leading to antiferromagnetic coupling (superexchange). Itinerant $d_{XZ}/d_{YZ}$ electrons, on the other hand, don't have this limitation and thus have ferromagnetic Hund-coupling with the local moments. The magnetic order depends on the relative density of localized moments and itinerant electrons, and on the orbital nature of the latter. If $d_{XY}$ itinerant electrons are occupied and their density is much lower than that of the localized moments, they will induce RKKY interactions that ferromagnetically align the moments (103). Dilute itinerant $d_{XZ}/d_{YZ}$ electrons would also induce FM order, but this time through a double-exchange mechanism (112, 113), driven by the Hund's rule alignment of itinerant electron spins and the spin of the localized moments. Due to this alignment the itinerant electrons can hop most easily between localized sites whose spins are FM aligned, and the resulting gain in kinetic energy leads to an effective FM interaction between the localized moments. If Rashba spin-orbit interactions are also present, the spin of the itinerant electrons would rotate as it hops between sites, stabilizing a spin spiral ground state (109). From general symmetry arguments it is possible to show that skyrmions and cone states could also be a stable ground state (111).

Further interesting phases may be found when the density of local moments becomes comparable to the itinerant electron density. Antiferromagnetic coupling to itinerant $d_{XY}$ electrons will lead to a non-magnetic Kondo screened phase (17, 30, 110). However, once the $d_{XZ}/d_{YZ}$ electrons, which have competing FM interactions with the moments, begin to accumulate, the Kondo phase will start to diminish and as a result the corresponding Kondo temperature will decrease as the carrier density is increased (30, 110). This scenario is so far the only model that can explain the experimentally observed reduction of the metamagnetic critical field with increasing carrier density. At still higher carrier densities, the random sign of the RKKY interactions mediated by the itinerant electrons will lead to a spin-glass state (110).

While a clarified, coherent view of magnetism at the LaAlO$_3$/SrTiO$_3$ interface begins to emerge, there still remain more open questions than answers – local magnetic structure, coupling to the electronic degrees of freedom, and the possibility for magnetism at ferroelastic domain walls are all areas that require further investigation. Ultimately, as our understanding of magnetism in the oxides further crystallizes, we may hope to exploit the gate-tunable phase diagram to control magnetic structures and interactions at the nanoscale.

## 8. Ferroelasticity – the microscopic structure of domains

Until recently, studies on complex oxide heterostructures have focused primarily on the electrical and magnetic properties of the conducting interface, neglecting the influence of the rich structural phenomenology of the $SrTiO_3$ substrate, in particular its ferroelasticity. The primary reason experimental progress has been hindered is that standard surface science probes such as STM are insufficient for studying the buried interfaces presented by oxide heterostructures. In this section, after first extending the discussion of ferroelasticity in $SrTiO_3$, we describe the results of experiments employing novel nanoscale probes to study the ferroelastic state in $SrTiO_3$ and its important influence on the $LaAlO_3/SrTiO_3$ interface.

### 8.1 Ferroelastic domains in $SrTiO_3$

The ferroelastic state of $SrTiO_3$ emerges below $T = 105$ K, characterized by the breaking of the cubic point-group symmetry of the unit cell (7, 8, 114). Below this transition temperature, the crystal unit cells become rectangular prisms, whose long axes may be oriented along any of the crystal axes, $X$, $Y$, or $Z$ (Fig. 7c). By analogy with ferromagnetism or ferroelectricity, the transition to the ferroelastic state is accompanied by the formation of domains characterized by the orientation of their unit cells which distribute to minimize the built-in strain profile in the bulk $SrTiO_3$. These domains obey a simple set of tiling rules: to minimize dislocations, unit cells in neighboring domains with orthogonally elongated unit cells must share their a-axes at the domain walls (twin boundaries). The resulting domain wall patterns are outlined in Figure 7c. Such ferroelastic domain wall patterns have been observed in bulk $SrTiO_3$ via polarized optical microscopy more than half a century ago (9, 10).

### 8.2 Domain structure and influence on transport

With the advent of novel probes suited for studying buried interfaces, experiments are now revealing the important consequences of the ferroelastic domain structure for the 2DES at the interface between $LaAlO_3$ and $SrTiO_3$. Advances in scanning single electron transistors (SETs) now allow for non-invasive imaging of the local mechanical response and electrostatic landscape in these buried interfaces with unprecedented sensitivity by using a nanotube quantum dot as a charge detector placed at the end of a scanned-probe cantilever (Figure 7b, upper panel) (4). Recent scanning SET studies of $LaAlO_3/SrTiO_3$ have uncovered that an applied back gate voltage couples to the domains, possibly through anisotropic electrostriction (115) or charged/polar domain walls (116-118), and can readily move the domain walls. This gate voltage-induced domain wall motion leads to an anomalously large piezoresponse in $SrTiO_3$ (4, 119). By taking into account the above-mentioned tiling rules, these measurements of the local electromechanical response (Figure 7b, lower panel) even allow for the specific labeling (Figure 7d) of the domain orientations as viewed from above the $LaAlO_3/SrTiO_3$ interface.

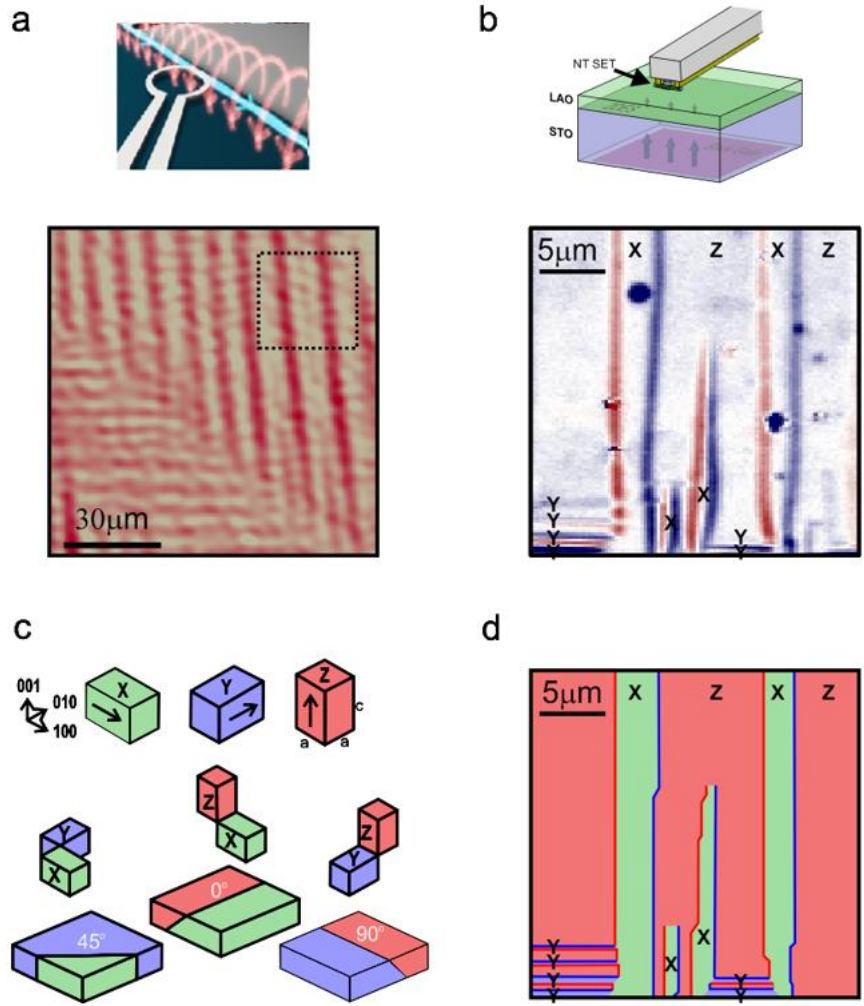

**Figure 7: Ferroelasticity**. (a) Illustration of a scanning squid probe used for imaging current flow is shown in the upper panel. Magnetic field lines (red) generated by a current path (blue) are seen threading the squid loop. An image of current flow through the LaAlO$_3$/SrTiO$_3$ interface captured with such a scanning squid probe is shown in the lower panel. Red coloring indicates high current density, while white indicates low density. Strikingly, the current is observed to flow in narrow channels. (Reprinted with permission from Beena Kalisky.) (b) Scanning nanotube-based single electron transistor (SET) imaging. Illustration of a nanotube-based SET probe hovering above an LaAlO$_3$/SrTiO$_3$. A nanotube is shown in the upper panel. By scanning this SET across the LaAlO$_3$/SrTiO$_3$ surface, the local electrostatic landscape can be imaged. The lower panel shows an image taken via scanning SET of gate voltage-induced domain wall motion in LaAlO$_3$/SrTiO$_3$. The image contains various patterns of stripes, each representing the motion of either the rising (red) or falling (blue) edge of a potential step in the surface induced by an oscillating back gate voltage. (Adapted by permission from Macmillan Publishers Ltd: Reference (4), copyright 2013.) (c) Tetragonal domain tiling rules. Below the ferroelastic transition, the cubic SrTiO$_3$ unit cells distort into elongated rectangular prisms, forming domains characterized by the orientation of their long $c$-axes that may lie along the $X$ (green), $Y$ (blue), or $Z$ (red) directions. Intersections of different domain orientations must share their short $a$-axis to minimize dislocations, forming twin boundaries with characteristic angles of either 0°, 45°, or 90°. (Adapted by permission from Macmillan Publishers Ltd: Reference (4), copyright 2013.) (d) Schematic map labelling the domain orientations imaged in panel (b). The dark red and blue stripes represent the domain wall boundaries. The coloring of the domains follows from the tiling rules outlined in panel (c).

Scanning SET measurements have also shown that the ferroelastic domains in SrTiO$_3$ create a varying potential landscape at the LaAlO$_3$/SrTiO$_3$ interface (4) , where the potential differs between regions of in-plane (X or Y) and out-of-plane (Z) domains. To maintain electrochemical equilibrium, charge must transfer between domains of varying potential, creating a dipole-like charge distribution at the domain walls, possibly leading to strong enhancement of charge density in the 2DEG at these walls. As electron mobility is a strong function of carrier density in LaAlO$_3$/SrTiO$_3$ (120), such charge enhancement could then lead to channeled current flow (121).

Such channeled current flow has in fact recently been observed in scanning squid measurements (3) (Figure 7a). In these measurements, the magnetic field generated by current flow is imaged, and this data is used to generate a spatial map of current density at the LaAlO$_3$/SrTiO$_3$ interface. By comparing current maps recorded at sample temperatures above and below the ferroelastic transition temperature for SrTiO$_3$, these measurements discovered that current within the 2DEG flows in narrow parallel channels that arise from the ferroelastic domains in the underlying SrTiO$_3$. At this stage, the experimental resolution does not yet allow for distinguishing whether the current flows preferentially within the bulk of one domain orientation, or instead actually flows along the domain walls. Furthering the understanding of the physics of this channeled flow promises to be a major thrust of future research in the oxide interfaces.

## 8.3 Toward domain-based nanosystems

As with the other emergent properties of SrTiO$_3$ previously discussed, ferroelastic domains increase the complexity of an already very rich class of physical systems, making interpreting macroscopic experiments more challenging. However, this new degree of freedom, if controlled and probed on the nanoscale, in fact presents exciting new opportunities for engineering unique low-dimensional nanostructures (122). The domain distribution can be manipulated via applied strain as well as with gate voltages, arming researchers with a new level of experimental control over transport in LaAlO$_3$/SrTiO$_3$ interfaces. For example, one can envision exploiting the domains to create one-dimensional conducting channels, which will inherit all the intriguing properties of the 2DEG discussed above, such as superconductivity, magnetism, and strong spin orbit coupling. These domain-based nanostructures may be combined with other novel nanoscale patterning techniques, including the AFM-based approaches (65) which are the focus of the next section.

## 9. Functional Devices down to the Nanoscale

Having discussed the novel physical phenomena and relevant nanofabrication techniques of LaAlO$_3$/SrTiO$_3$, we now survey progress toward creating functional devices from this material system, with feature sizes ranging from the microscale to the nanoscale.

## 9.1 Microscale

By exploiting the unique properties of conducting complex oxides interfaces, one can envision creating oxide-based circuit elements that feature novel switching mechanisms relying on the metal-insulator transition or strong electron interactions. Building on advancements made toward realizing such oxide electronics using zinc- and vanadium-based materials (123, 124), exciting progress has been made in developing LaAlO$_3$/SrTiO$_3$-based circuits. Capacitors (125) and diodes (126) were among the first functional circuit elements made from LaAlO$_3$/SrTiO$_3$. More recently, fully integrated circuits have been created (Figure 8a) from top-gated, microscale LaAlO$_3$/SrTiO$_3$ field-effect transistors, including arrays of ring oscillators composed of several inverter stages (127). In addition to the active LaAlO$_3$/SrTiO$_3$ transistors, these devices also featured oxide-based resistors and interconnects, demonstrating the potential for all-oxide circuitry. Increasing the complexity and further miniaturizing these circuits promises to be an important continued research direction.

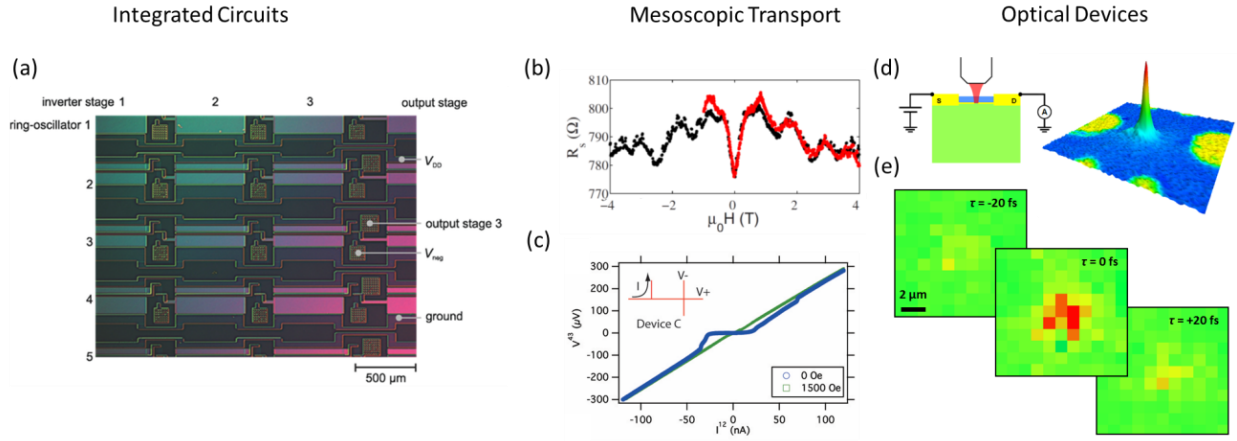

**Figure 8: Functional Devices down to the nanoscale.** (a) Optical microscopy image of a series ring oscillators made from LaAlO$_3$/SrTiO$_3$, demonstrating the ability to fabricate fully-integrated circuits with micron-sized features from oxide interfaces. **(Permission)** (b) Universal conductance fluctuations. Quantum oscillations in the conductance as a function of magnetic field are observed in LaAlO$_3$/SrTiO$_3$ devices with dimensions smaller than the phase coherence length. The data shown are from a Hall bar device of size 4 μm × 1.5 μm. The red and black coloring of the data points denotes two separate magnetic field sweeps, highlighting the reproducible pattern of fluctuations. (Adapted with permission from Reference (130). Copyright 2010 by the American Physical Society.) (c) Nonlocal resistance in AFM-sketched nanostructures. The inset schematic shows the current path $I$ and terminals for voltage measurement $V^+$ and $V^-$ in a device of length 10 μm and width 10 nm. Cooper pairs form at sufficiently low temperature, blocking spin transport without current flow, and thus suppressing nonlocal voltage below the critical current (blue curve). Application of an external magnetic field suppresses the superconductivity and restores the nonlocal response (green curve). The data shown were taken at $T = 65$ mK. (Reprinted with permission from Reference (6). Copyright 2013 by the American Physical Society.) (d) Nanoscale devices fabricated by c-AFM can be switched optically. (e) Ultrafast photoresponse of c-AFM LaAlO$_3$/SrTiO$_3$ with femtosecond dynamics.

## 9.2 Mesoscale

An important series of developments toward the creation of functional LaAlO$_3$/SrTiO$_3$ nanodevices is the realization of *mesoscale* devices. For these devices, the physical size of the device itself influences the electronic properties due to the onset of quantum effects. One indication of such mesoscopic transport is the appearance of universal conductance fluctuations (UCF) (128, 129), signified by fluctuations in conductance on the order of $e^2/h$, the quantum of conductance. UCF was observed in LaAlO$_3$/SrTiO$_3$ samples (Figure 8b), which were small enough with respect to the electrons' phase coherence length to avoid averaging away these quantum oscillations (130). Lithographically-defined structures with channels as narrow as 500nm have exhibited in addition to UCF also superconductivity (62), which might be more sensitive to fabrication imperfections, indicating that such fabrication techniques do not deteriorate device quality. In micron-scale devices fabricated by reactive-ion-etching of LaAlO$_3$, an anomalously large capacitive gating effect with strong dependence on device size has been observed (131), demonstrating that new and useful features can emerge with shrinking size.

## 9.3 Nanoscale

Conductive-AFM lithography has proven to be a powerful tool in creating nanoscale devices in LaAlO$_3$/SrTiO$_3$. Devices with rather complicated geometries can be made by directly writing their pattern with the c-AFM tip. The simplest devices made are nanowires, which give room temperature resistances of $\sim 100$ kOhm/μm, and can reach anomalously-high room temperature mobility $\sim 400$ cm$^2$/Vs, as their width is decreased (132), rivaling the best silicon-based nanowires. By ramping the voltage applied to the c-AFM tip while moving across the LaAlO$_3$ surface, diodes with positive and negative breakdown voltages can be formed (133). This ability to shape the symmetry of the confining potential that defines the nanowire can potentially be used to modulate spin-orbit interactions in space, useful for quantum information processing (134). A simple tunnel junction is formed by negatively biasing the c-AFM tip to locally restore the insulating state of the LaAlO$_3$/SrTiO$_3$ interface. Such junction has useful device properties by itself, having nonlinear $IV$ characteristics governed by the activation barrier and the nanowire resistance (135).

Low temperature transport properties of LaAlO$_3$/SrTiO$_3$-based nanowires exhibit many remarkable and unexplained effects. Similar to the 2D interface, nanowires undergo a superconducting transition below 200 mK (105). Nonlocal transport is observed in both normal (6) and superconducting (136) states (Figure 8c). In the superconducting state, this effect is attributed to charge imbalance (137, 138), while in the normal state the situation is less clear. The combination of ferromagnetism, strong spin-orbit coupling, and quasi-one dimensionality may lead to helical transport (104), an important starting component for creating Majorana fermions (139, 140), though other explanations may be possible.

A more complex, voltage-tunable nanoscale junction can be created by adding a third electrode (141). These devices can be operated as transistors with on/off ratios up to 10,000 and are functional up to the GHz frequency range (142). To date, $p$-type transistors have not yet been produced with this c-AFM technique, preventing the production of complementary (CMOS-like) devices. Nevertheless, unipolar architectures of many common circuit elements are in principle possible (*e.g.* NMOS logic). Transistor-like behavior has even been achieved using an optical gate (143). These LaAlO$_3$/SrTiO$_3$ photodetector gates can be operated using ultrafast sources, allowing generation and detection of broadband THz emission at the nanoscale (Figure 8d,e) (144). This class of devices may find applications in local molecular characterization and sensing.

An interesting extension deeper into the nanoscale regime is the creation of single-electron transistors (SETs) (145). The ability to create nanoscale islands with tunable couplings represents a fundamental scaling limit for solid-state device fabrication. Single electron transistors fabricated from LaAlO$_3$/SrTiO$_3$ using c-AFM can be scaled to less than 2nm in size, where they exhibit a ferroelectric-like bistable state and show Coulomb blockade up to $T = 40$ K (5). As the number of electrons hosted by the SET can be precisely controlled (0, 1, 2, etc...) these devices may serve as potential building blocks for solid state quantum simulation.

# 10. Future Directions

Over the course of this review, we have emphasized the unique advantages of complex oxide interfaces as platforms for exploring highly-varied physical phenomena. Within the same interface, one may simultaneously find a rich set of fundamental physics phenomena. However, the high level of disorder present so far in these materials hinders the ability to disentangle and control these various effects independently, thus limiting this system from reaching its full potential.

One clear route for overcoming the disorder is a continued improvement of material quality. Through improved design and growth of these interfaces, it should be possible to separately control the different phenomena, and ideally switch them at will, for example with externally applied voltages. Indeed, the material science mastery of these interfaces is still in its infancy as compared with the much more mature field of semiconductors. Many of the techniques used to advance the field of semiconductors, such as the complete separation of dopants from the conducting channel, the lattice matching between different materials, strain engineering, and the optimization of MBE growth (146), if successfully adopted could similarly revolutionize the field of oxide heterointerfaces.

## 10.1 Beating complexity by probing on the nanoscale

Faced with this complex, disordered landscape of coexisting phenomena, there exists another approach using nanoscale probes to circumvent the elements that currently evade control. Nanoscale probes have already shown that within a single $LaAlO_3/SrTiO_3$ sample, different phenomena like ferroelastic domain walls, superconductivity and ferromagnetism can coexist on micrometer scales. When probed on macroscopic scales, the underlying physics of these competing phases is washed out by spatial averaging. However, by probing the interfaces on scales smaller than typical scales over which these phenomena vary, they can be independently studied. In fact, the natural tendency of these different phenomena to coexist can even be exploited to realize novel nanosystems that are presently too difficult to intentionally engineer but can be readily studied with existing nanoscale probes. One such serendipitous arrangement, for example, could be the creation of one-dimensional conducting channels at ferroelastic domain walls.

Furthermore, existing nanoscale probes can measure important physical quantities of the oxide interfaces that have so far been inaccessible via macroscopic approaches. Scanning single electron transistors may be used for example to spatially map the full subband structure, to directly identify the dopants that give rise to the itinerant electrons, and to study the nature of electron localization at the interface. Magnetic force microscopes and the new generation of nanoscale scanning SQUIDs (147) could be used to resolve the physics at ferroelastic domain walls. Optical (148) and microwave (149) nanoscale probes that have proved useful in analogous materials systems are likely to also discover new phenomena in these interfaces.

## 10.2 Hybrid systems

Although complex oxide heterointerfaces are buried, they are quite close to the surface as compared with semiconductor heterostructures. Due to this proximity, one can envision engineering unique couplings between oxide interfaces and other two- or even one-dimensional materials outside of the complex oxides family, such as $MoS_2$, ferromagnetic thin films, high-temperature superconductors, graphene, nanotubes and nanowires. Functional electronic, magnetic, and structural couplings, as well as direct tunneling may be explored, extending the parameter space within which to search for novel phenomena. While it is difficult to forecast what discoveries will come from these hybrid systems, their high degree of flexibility makes them a very intriguing future research direction.

## 10.3 Engineered physics with d-orbital electrons on the nanoscale

Once a higher degree of cleanliness and control within the complex oxides has been achieved, it will be possible to design systems which maximally exploit the rich properties endowed by the $d$-orbital electrons in the oxides. Independent control over the constituent properties will make the oxide interfaces a powerful laboratory for engineering correlated electron physics on the nanoscale in new, previously inaccessible parameter regimes, including Kondo lattices, engineered Mott insulators, unconventional realizations of the quantum Hall effect, superconductivity in 1D channels, and Majorana fermion states. As both material quality and capabilities of nanoscale probes continue to advance, we foresee that the years ahead will be quite exciting for researchers in the field of complex oxide heterostructures.

## Acknowledgements

We would like to thank E. Altman, J. Ruhman, and K. Michaeli for critical reading of the manuscript. S.I. acknowledges financial support by the Israel Science Foundation (No. 1267/12), the Minerva Foundation and the ERC Starting Grant (QUANT-DES-CNT, No. 258753), and the Marie Curie People Grant (IRG, No. 239322). P.I. is grateful for financial support from the Air Force Office of Scientific Research (FA9550-12-1-0268). J.L. acknowledges financial support from the National Science Foundation (DMR-1104191, DMR-1124131), Army Research Office (W911NF-08-1-0317), and Air Force Office for Scientific Research (FA9550-10-1-0524, FA9550-12-1-0057, FA9550-12-1-0268), and the Office of Naval Research (N00014-13-1-0806).